# An approach of local band average for the temperature dependence of lattice thermal expansion

Mingxia Gu,[a] Chang Q Sun[a,*] and Yichun Zhou[b,*]

[a] School of Electronic and Electrical Engineering, Nanyang Technological University of Singapore, 639798

[b] Key laboratory of low-Dimensional Materials and Application Technologies (Xiangtan University), Ministry of Education, Hunan 411105, China

Abstract

It has long been puzzling regarding the mechanism behind the nonlinearity of lattice thermal expansion at low temperatures despite modeling considerations from various perspectives in classical or quantum approximations. An analytical solution in terms of local bond average is presented herewith showing that the thermal expansion coefficient follows closely the specific heat of Debye approximation without the involvement of mode Grüneisen constant or the bulk modulus. Matching predictions to experimental observations using the Debye temperature and the atomic cohesive energy as input evidences that the current approach may represent the true situation of temperature induced lattice expansion though the exact form of phonon density of states need to be considered for further refinement.

[*] Electronic Address: ecqsun@ntu.edu.sg; zhouyc@xtu.edu.cn;

I    Introduction

Thermally-induced lattice expansion of a specimen and the thermal expansion coefficient (TEC or $\alpha(t)$) are fundamentally of great importance to the performance of materials for devices or engineering constructions, which have long been studied both experimentally[1,2,3,4] and theoretically.[5,6,7,8] For example, when a sample layer is grown on a substrate, problems may arise from different TECs (or the thermal mismatch). During deposition, residual stresses can be built up at the interface between the sample layer and the substrate because of the mismatch between the TECs of these layers.[9,10] Likewise, the thermal stress can also develop in the sample during annealing,[11] and hence affects the performance of the device.[12] The TECs of different materials have been well studied experimentally during the past decades; however, even for the same material such as diamond,[2,3,13,14] silicon,[4,15,16,17] and GaN,[1,18,19,20] the available experimental data scattered significantly depending on the measuring techniques.

From microscopic point of view, the temperature dependence of the bond relaxation is usually attributed to the anharmonicity of the interatomic potential. The $T$-dependent TECs are found analogous to the temperature dependent specific heat,[7,18,21] and they have been described by a number of sophisticated models [1,2,4,6,21,22,23,24] from the perspective of classical thermodynamics and lattice quantum vibrations. Numerically, all the models could reproduce the general trends of measured temperature dependent TECs with a number of adjustable parameters needing physical indication. As the temperature dependent bond length $l$ could be well fitted by a polynomial $\sum_0^4 A_n T^n$, and hence the TECs were also able to be described by a polynomial empirically.[2,21,22] The observed TECs sometimes could also be well fitted using exponential terms.[17,23] The *ab initio* method, which computed the total energy for the equilibrium and distorted atomic configuration, was also used to calculate the temperature dependent TECs for Al and W,[24] with the involvement of bulk modulus, mode Grüneisen constant, and the concave parameter as adjustable variables. Phenomenological lattice dynamical theory in quasiharmonic approximation is also well applied to describe the temperature dependence of TECs.[7] However, consistent understanding of the atomistic origin of the thermal expansion and a unified form of expression is yet a challenge. The objective of this work is to show that a simple and straightforward analytical solution can be developed from the perspective of local bond average (LBA) and the TEC follows closely the temperature dependence of the specific heat.

II      Principle: local bond average

A bulk solid is formed by numerous atoms with bonds connected one to another. For a given specimen, no mater whether it is crystal, non-crystal, or with defects or impurities, the nature and the total number of bonds do not change under the external stimulus of temperature unless phase transition occurs. However, the length and strength of all involved bonds will response to the operation temperature. If the functional dependence of a detectable quantity on the bonding identities (bond length, bond strength and bond nature) is known, one is able to predict the performance of the solid by focusing on the response of bond length and energy to the specified stimulus. Therefore, the performance of the representative bonds can represent the specific sites or the average of the representative bonds for the entire sample. This LBA approach may represent the situations of measurements and theoretical computations that collect statistic information from large number of atoms of the given specimen. Furthermore, compared with the measurement and computation, the LBA could discriminate the behavior of local bonds at different sites.

Generally, the material dimension expands upon temperature increase. The bond length of a specimen has the following relation with respect to the temperature under consideration,

$$l = l_0 \left[1 + \int_0^T \alpha(t) dt \right] \qquad (1)$$

where $\alpha(t)$ is the thermal expansion coefficient. From bonding energetics and the LBA view point, and by introducing the concept of interatomic potential, $u(r)$, $\alpha(t)$ follows

$$\alpha(t) = \frac{1}{l_0}\left(\frac{dl}{dt}\right) = \frac{1}{l_0}\left(\frac{dl}{du}\right)\left(\frac{du}{dt}\right) = \frac{C_v(T/\theta_d)}{-l_0 F(r)} \qquad (2)$$

because $\dfrac{du}{dl} = -F(r) > 0$ and $\dfrac{du}{dt} = C_v(T/\theta_D)$

where $u(r)$ is the pairing potential and $F$ represents the restoring force at a non-equilibrium point, $r \neq r_0$. Taking the Lennard-Jones potential, for example, The $-F(r)$ takes a value from 0 to a positive value of finite small. Since the thermally-expended bond deviates from the equilibrium position by a maximal amount of 5% of the equilibrium length at the melting point,[25] the force $F(r)$ depends non-linearly on the atomic distance $r$ if the anharmonicity is considered. $C_v$ is the specific heat per atom, which is assumed to follow Debye model, $C_v(t) = \dfrac{9R(t/\theta_D)^3}{N_A} \int_0^{\theta_D/t} \dfrac{x^4 \exp(x)}{[\exp(x)-1]^2} dx$ or Einstein model. The $\theta_D$ and $N_A$ are Debye temperature and the Avogadro constant, respectively. However, Ref 24 (figure 9 and 10) demonstrated insignificant difference in numerical between the two models in describing the trend of $C_v$ for Al and W. Both the Einstein model and the Debye model match each other very well at high temperature. Because the Debye model could fit better to the measurement at low temperature, the Debye approximation will be used in subsequent discussions. The integration of the specific heat represents the inner energy rising due to thermally excited lattice vibration in all possible modes. It is not practical for one to discriminate the acoustic from the optical modes or the anharmonic from the anharmonic. For the simple lattice structure, only acoustic modes exist but for compounds and diamond structures, both optic and acoustic modes are involved.

By considering the fact that the product of bond length at 0 K ($l_0$) and the force $F(r)$ is in the dimension of atomic bonding energy, we have $l_0 F(r) = A_1(r) E_B(0)$, where $E_B(0)$ is the intrinsic atomic bonding energy at 0 K, and $A_1(r)$ is a r-dependent coefficient. The $T$-dependent TEC can be rewritten as,

$$\alpha(t) = \frac{C_v(t)}{l_0 F(r)} \cong \frac{C_v(t)}{A_1(r) E_B(0)} = A(r) C_v(t) \tag{3}$$

Hence, $A(r) = [-l_0 F(r)]^{-1} = [A_1(r) E_B(0)]^{-1}$ and $A(r)$ is related to the restoring force at non-equilibrium position $r$, $F(r)$. Generally, the change of lattice constant or the bond length with increasing temperature is in a range of $10^{-6}$. Taking Lennard-Jones (LJ) potential for example, within this small range the restoring force $F(r)$ is in an order of $10^{-5}\varepsilon$, where $\varepsilon$ (several eV) is the magnitude of depth of the potential well in LJ potential. Therefore, $F(r)$ can be approximated to be a constant, and hence the $A(r)$. From Eq. (3), the parameter $A$ and the Debye temperature are the only two fitting parameters. For the performance of the single bond, the change of pressure is not considered in the approximation such that the thermally-induced expansion is assumed to be under constant pressure. The $C_v$ should add the P-constant term causing an offset of the curve linearly, which should compensate for the F(r). From the harmonic approximation F® depends linearly on temperature. The anharmonic contribution at the small strain is in significant. Eqs (2) and (3) justify that the TEC follows closely the trend of the specific heat, which is numerically consistent with most of the existing models.[7,8,21] Furthermore, if the Einstein's model is applied, the exponential terms in Reeber's model[1] can be obtained. If the exponential term is expanded in Tayor's serious, the temperature dependent polynomial TECs is quite obvious, which also elaborates some empirical models.[21,22] Therefore, with the current justification, the previous models discussed[1-22] are numerically correct.

Following Eqs. (1) and (3), the bond length $l$ can be expressed as

$$\frac{l}{l_0} = 1 + \int_0^T \alpha(t)dt = 1 + A \times \int_0^T [C_v(t) + \delta]dt = 1 + AU(T)$$

$$\text{where } U(T) = 9RT(T/\theta_D)^3 \int_0^{\theta_D/T} \frac{x^3}{\exp(x)-1} dx$$

(4)

which justifies the relations proposed by Roder[18].

We have thus derived the relations for the temperature dependence of TEC and thermally induced bond length expansion from the perspective of LBA approximation. The constant $A$, in unit of (eV/atom)$^{-1}$, being related to the bond energy at equilibrium bond length determines the magnitude of the TECs at higher temperature; the Debye temperature, $\theta_D$, determines the width of the shoulder at low temperature in the $\alpha(t)$ curve.

III    Results and discussion

In order to reproduce the experimentally measured TECs, first of all, the linear dependence of $a(T)$ at high temperature is used to estimate the value of $A$. The $A$ is then used as an initial input in the subsequent fitting iteration. By careful reproduction of the available experimental data, the $A$ and $\theta_D$ can be obtained as tabulated in Table I for the considered samples. Using the obtained $A$ and $\theta_D$ we can also fit the temperature dependent lattice parameter using Eq. (4). A fine tuning of $A$ and $\theta_D$ is necessary because of the difference in the source of data and errors in the measurement. The refined $A$ and $\theta_D$ are also given in Table I for comparison. The current $\theta_D$ values derived from fitting the $T$-dependent lattice parameter and TECs are consistent with Redor's result.[18] As the fitting parameter $A$ relates to $E_B(0)$ that can be obtained from fitting to temperature dependent Young's modulus and Raman optical phonon shift.[26,27] With the given atomic bonding energy $E_B(0)$, the amplitude $A_1$ can readily be obtained.

Figure 1 compares the fitting to the measurement temperature dependence of TECs using Eq (3). It shows that the current approach covers the general trend for $\alpha(T)$. Exceedingly well agreement with the measured data has been obtained for nitrides, like AlN, $Si_3N_4$, and GaN. However, the current model does not reproduce the observed negative TECs in group IV elements. Generally, most materials expand upon heating, although, very rare, some materials expand upon cooling. The unusual behavior of materials having negative TECs for some tetrahedral materials have been considered arising from the negative Grüneisen parameters of the transverse acoustic phonons near the Brillouin-zone boundary.[5,6,28] Further more, the model predictions match better the lattice behavior at low temperatures than at high temperatures for some pure metals, such as Au, Cu, and Al showing that the TECs continue increasing with temperature. In metals, in addition to the phonon contribution to thermal expansion, free conduction electrons also play a role in temperature dependent lattice constant change.[29] For some ferromagnetic metals, like Ni and Fe, the measured TECs exhibit a phase transition from ferromagnetic to paramagnetic and the abrupt feature may arise from spin contribution to the specific heat.[30] It is not surprising that these unexpected features are beyond the scope of the current model because we used an ideal case of the phonon density of states derived from long wavelength at the Brillion zone center and only the phonon contribution is considered. The Debye approximation of specific heat assumes that the phonon density of states in an elastic medium is ideally proportional to $\omega^2$. In reality, one has to consider the exact form of the phonon density of states that are hardly available experimentally or theoretically.[5] Nevertheless, the phonon contribution to the thermal properties is dominant and a precise prediction of the *T*-dependent TECs can be made if the exact density of states $g(\omega)$ vs $\omega$ is used. Anyway, well fitting to experimental available TECs reveals that the current model may represent the dominant contribution from bonding energetic point of view.

Figure 2 shows that the current approach allows us to reproduce the *T*-dependent lattice parameters of various materials. The fact that much better match of the lattice constants compared to the match of the TECs for the same elements indicate that experimental errors can not be ignored in practice.
The thermally induced energy rise arise from lattice vibration of all possible modes. For pure metals, only accoustic modes contribute. Hoewever, both optical and acoustic modes are activated. It is there fore not practical for one attempt to discriminate contribution from acurstic fornm optic mode the the thermal energy increasement. On the other hand, both Debye and Einstein approximations for the specific heat are equally important and provide insignificant difference in numerical…..
A value discussion.???. constant justification?

IV Conclusion
It is found that the temperature dependent TECs follws follows closely the general trend of the temperature dependent specific heat. The slope in the high temperature range relates to the bond energy and the width of the shoulder relates to Debye temperature. No other parameters such as the bulk modulus and the Grüneisen parameter are needed in the current LBA approach. Exceedingly well agreement to the measured TECs for nitride and the general trends for metals and diamond structures may evidence the LBA approach that may represent the true situation of observations though refinement can be made by using the real phonon density of states for a particular specimen.

This project is supported by MOE (RG14/06), Singapore and Chinese NSF (10525211 and 10772157).

Table and Figure Caption

Table I Parameters derived from fitting to the TECs and lattice parameters. The documented Debye temperature is obtained from Ref 31.

Fig 1 Examples of Lennard-Jones potential and the force as a function of lattice distance $r/r_0$, where $r_0$ is the distance at equilibrium.

Fig 2 Comparison of the predictions (curves) to the measured (scattered) temperature dependence of the TECs of (a) AlN, (Ref 21) $Si_3N_4$ (Ref 21), and GaN (Ref 1, 32); (b) Si (Ref 4, 6, 16,17) , Ge (Ref 33, 34) and Diamond (Ref 1, 2, 3, 6, 13, 14); (c) Au, Cu, and Al (Ref 35); (d) Ni and Fe (Ref 35).

Fig 3 Comparison of the predictions (curves) to the measured (scattered) temperature dependence of the lattice parameter expansion for (a) AlN (Ref 13), and GaN (Ref 1, 19,20); (b) Si (Ref 36), Ge (Ref 33) and Diamond (Ref 2, 14); (c) Au, Cu, and Al (Ref 35); (d) Ni and Fe (Ref 35).

Table I

|  |  | Si | Ge | C | AlN | | Si$_3$N$_4$ | GaN | | Au | Cu | Al | Ni | Fe |
|---|---|---|---|---|---|---|---|---|---|---|---|---|---|---|
| Reference | $\theta_D$ (K) | 647 | 360 | 1860 | 1150 | | 1150 | 600 | | 170 | 315 | 420 | 375 | 460 |
| $\alpha$(t) | $\theta_D$ (K) | 1000 | 600 | 2500 | 1500 | | 1600 | 850 | | 400 | 500 | 450 | 600 | 600 |
|  | A | 0.579 | 0.966 | 0.811 | 0.888 | | 0.502 | 0.637 | | 2.241 | 2.588 | 3.322 | 2.009 | 1.777 |
| $l$(t) | $\theta_D$ (K) | 1100 | 500 | 2150 | 1500 | | 1400 | 800 | | 400 | 500 | 500 | 600 | 600 |
|  | A | 0.579 | 1.035 | 0.792 | 0.946[a] | 0.811[b] | 0.888 | 0.637[a] | 0.618[b] | 2.105 | 2.588 | 3.554 | 2.144 | 20.09 |
|  | $l_0$ (Å) | 5.4286 | 5.65 | 3.5661 | 3.1095 | 4.9774 | 7.7335 | 3.1893 | 5.1830 | 4.07 | 3.6 | 4.036 | 3.513 | 2.82 |
| Mean | $\theta_D$ (K) | 1050 | 550 | 2325 | 1500 | | 1500 | 825 | | 400 | 500 | 475 | 600 | 600 |
|  | A | 0.579 | 1.001 | 0.802 | 0.882 | | 0.695 | 0.631 | | 2.173 | 2.588 | 3.438 | 2.076 | 10.93 |

[a] a-axis

[b] c-axis

Figures

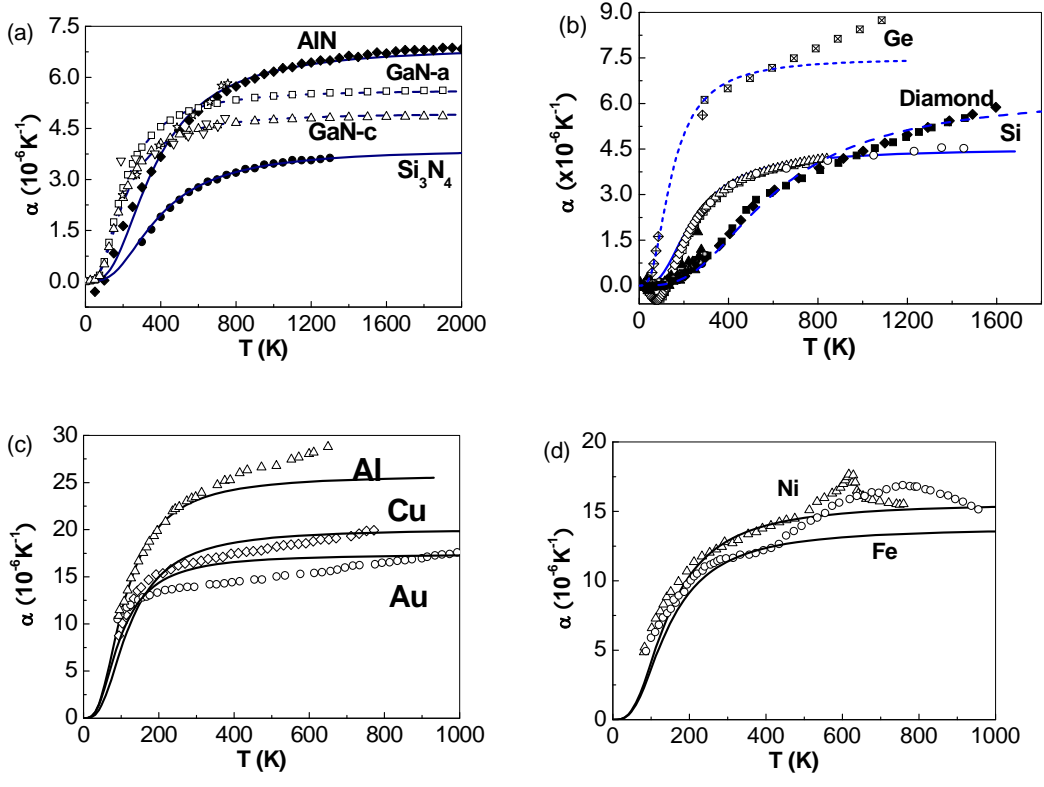

Figure 1

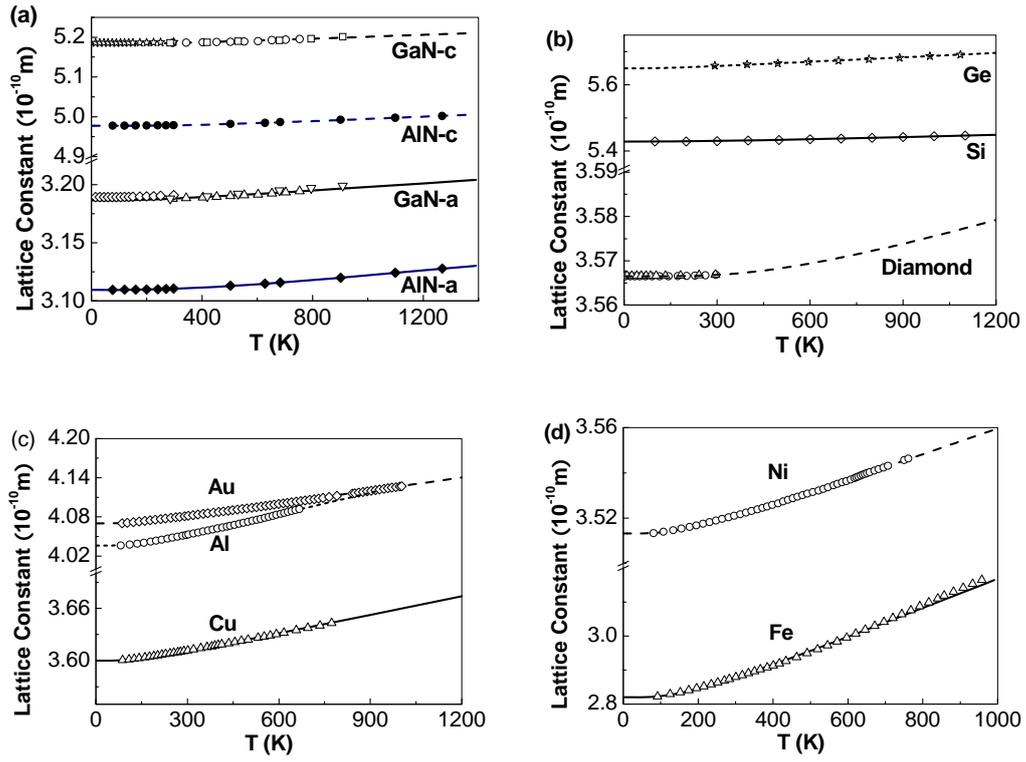

Figure 2